\documentstyle[12pt]{article}
\begin{document}
\begin{center}
\begin{large}
Some Applications of the Difference Analysis  
for Stochastic Systems \\ [0.3cm]
\end{large}
A. Yu. Shahverdian (1) and A. V. Apkarian (2)  \\ [0.2cm] 
(1) Yerevan Physics Institute, Yerevan, Armenia \\ 
(2) Northwestern University, The Medical School, Chicago, IL, U.S.A. \\
\end{center}
\begin{abstract}
The work relates to a new way for analysis of one-dimensional 
stochastic systems, based on consideration of its higher order 
difference structure. From this point of view, the deterministic 
and random 
processes are analyzed. A new numerical characteristic 
for one-dimensional 
stochastic systems is introduced. The applications to single 
neuron models and neural networks are given.
\end{abstract}

\section{Introduction}

This paper presents some applications of the difference analysis, 
has been suggested in authors' recent works [1-3]. 
The approach has been detected on computational study [1] of neural activity:
we observed that sequences of higher order absolute 
differences, taken from periodically stimulated neuron's
spike train, contain long samples on which the changes
in monotony (increase/decrease) are periodic.

The next Section 2 generalizes this observation and introduces a 
new characteristic for stochastic one-dimensional systems,
expressing this type of non-explicit periodicity in a
quantitative form. This notion, the numerical characteristic 
$\gamma$, being a measure of some minimal periodicity,
can also be treated as a new measure of irregularity. On 
instances of some nonlinear maps its numerical
comparisons with Lyapunov exponent are given.

The difference approach provides a strict
algorithmic base for detection of small chaotic fluctuations
at unstable bifurcation points. This is given for
logistic map, however the same analysis can be applied
for various systems, exhibiting the bifurcations.

The probabilistic systems permit rigorous study -- Section 3 
considers discrete random processes. The sequences 
of independent random variables as well as binary Markov 
processes are examined. 
In fact, we deal with a new type of limiting transition, taken
over the discrete random process. The applications to diffusion 
and Poisson processes are given.
We establish several theoretical results relating to
$\gamma$-characteristic of random discrete systems. 
We are based on Eggleston theorem [7] from ergodic theory
to describe the attractors for these random processes.

Section 4 considers stochastic models of single neuron
as well as establishes some probabilistic neuron's 
firing condition. 
We are again based on the difference analysis and use
a modified version of entropy. We claim that some 
new type of attractors, resulting in the approach suggested,
can be treated as a new extended memory in neural networks.
This appears to be well consistent with the brain theory
concepts of attractor computing and associative hierarchical
memory [8]. Finally, we prove a theorem, showing that if follow 
the difference approach, the noise in fact can be eliminated 
from arbitrary "noisy" neural network.

The paper is organized in such a way, that the most theoretical 
results can be easily deduced from the previous ones. The complete proofs of the other theoretical statements will appear 
in the version of this work submitted to publication to a regular journal.

\section{Deterministic processes}

\subsection{Finite differences and conjugate orbits}

The difference approach, suggested in [1-3], reduces the study of
stochastic properties of the orbits
${\bar X}=(x_{i})_{i=1}^{\infty}$, generated by given 
one-dimensional system, 
to analysis of alternations of the monotone increase or  
decrease of higher order absolute differences
$$ \Delta^{(s)}x_{i} =
|\Delta^{(s-1)}x_{i+1} - \Delta^{(s-1)}x_{i}|
\qquad (\Delta^{(0)}x_{i}= x_{i} \ ;
\ \ i,s = 1,2,3, \ldots) \ .
$$
In this section we describe some statements of the approach
and explain some basic notions involved in this work.

Let us have a one-dimensional stochastic system, generating numerical
sequences
${\bar X} = (x_{i})_{i=1}^{\infty}, \ 0\leq x_{i} \leq 1 $.
It is not difficult to see, that
for $1\leq s\leq k-1$ we have
\begin{equation}
 \Delta^{(s-1)}x_{i} = \mu_{k,s-1} +
 \sum_{p=1}^{i-1}(-1)^{\delta_{p}^{(s)}}\Delta^{(s)}x_{p} -
 \min_{0\leq i\leq k-s}(\sum_{p=1}^{i}
  (-1)^{\delta_{p}^{(s)}}\Delta^{(s)}x_{p})
\end{equation}   
where 
 $$     \delta^{(s)}_{p}= \left \{ \begin{array}{ll}
        0 & \Delta^{(s)}x_{p+1} \geq \Delta^{(s)}x_{p} \\
        1 & \Delta^{(s)}x_{p+1} <    \Delta^{(s)}x_{p}
	                          \end{array} \right.  
    \qquad  \  \mu_{k,s} =   \min \{ \Delta^{(s)}x_{i}:
 1\leq i \leq k-s\} \ ,
 $$
(it is assumed $\sum_{1}^{0} = 0$).
Using the recurrent formula (1) we transform the finite
orbit ${\bar X}_{k}$ into some special form, which
emphasizes its higher order difference structure. Namely,
we transform
${\bar{X}}_{k}=(x_{i})_{i=1}^{k}$
into the sequence ${\bar\zeta}_{k}$,
\begin{equation}
 {\bar \zeta}_{k} = ({\bar \lambda}_{k}, \  {\bar \mu}_{k}, \
 {\bar \rho}_{k})
\end{equation}
where 
\begin{equation}
 {\bar \lambda}_{k}
 = (\lambda_{1}, \lambda_{2},\ldots , \lambda_{k}), \qquad
 \lambda_{s}= 0.\delta_{1}^{(s)}\delta_{2}^{(s)} \ldots,
 \delta_{k-s}^{(s)} \qquad (s = 0, 1, \ldots, m)
\end{equation}
\begin{equation}
{\bar \mu}_{k} =   (\mu_{k,1}, \  \mu_{k,2}, \  \ldots ,
  \mu_{k,m}),    \qquad
{\bar\rho}_{k} =   (\rho_{k,1}, \  \rho_{k,2}, \  \ldots ,
\rho_{k,k-m})  \quad
   (\rho_{k,i} = \Delta_{i}^{(m)}) \ .
\end{equation}
The Eq.~(2) in fact represents the original orbit in a different
form -- one can see that applying the
recurrent procedure (1), the sequence
${\bar{X}}_{k}$ can be completely recovered by
${\bar{\zeta}}_{k}$. The difference method suggests to
study the sequences ${\bar\nu}=(\nu_{k})_{k=1}^{\infty}$
which terms are defined as follows:
\begin{equation}
 \nu_{k} =
 0.\delta_{1}^{(k)}\delta_{2}^{(k)}\delta_{3}^{(k)}\ldots 
\end{equation}
-- it is clear that
\begin{equation}
     |\nu_{k} - \lambda_{k}|\leq 2^{-k}
     \qquad (k\geq 1)   \ .
\end{equation}
The $\bar\nu$ is called the conjugate (to $\bar X$) orbit.
The approach distinguishes two cases --
continuous, when the quantities $\rho_{k,i}$ from (4) can take
arbitrary numerical values from interval $(0,1)$, and the 
discrete case,
when they take only a finite number of values.
In either case, given $\bar X$ we are interested in 
its higher order difference structure, that reflects the 
conjugate orbit $\bar\nu$.  

The computations performed in [1-2] show that for many
actual continuous-time systems the quantities
\begin{equation}
  ||{\bar\mu}_{k}||^{2} + ||{\bar\rho}_{k}||^{2}
    \qquad  \qquad (||(x_{1},\ldots, x_{m})|| =
    (\sum_{i=1}^{m}x_{i}^{2})^{1/2})
\end{equation}
as well as (due to relations (2) and (6)) the distances 
$||{\bar\zeta}_{k} - {\bar\nu}_{k}||$ converge to zero
with the exponential rate.
In contrary, the sequences ${\bar\nu}_{k}$ from (5)
mostly
have oscillating character and are attracted either to
an interval or to a thin set $\cal A$.
This means, that the dominant part of information, that
carries the original time series $\bar X$ and
which permits measurements, in fact is conveyed by its
conjugate orbit $\bar\nu$. For many
irregular systems the set $\cal A$ is the same for
different
orbits determining by different initial states. Hence, 
the $\cal A$ 
appears to be some (conjugate) attractor for the
system. Therefore, the main part of information, produced
by system during its evolution,
is contained in the attractor $\cal A$, which 
can be treated as a geometrical image of the whole
produced information.

The systems, generating some natural numbers belonging 
to a finite set, (e.g., as the outcomes in hazard games) 
should be classified to the discrete case. The
conjugate orbits are constructed by the same way: for instance,
if ${\bar X} = (x_{i})_{i=0}^{\infty}$ and $x_{i}\in \{0,1\}$,
then the difference sequences ${\bar X}^{(k)} =
(x_{i}^{(k)})_{i=0}^{\infty}$ are again the binary sequences and
the conjugate orbits consist of the terms
  $$\nu_{k} = \sum_{n=1}^{\infty}2^{-n}x_{n}^{(k)} \ . $$
However, for this case we do not have the convergence of (7) 
to zero as for 
continuous-time systems. Instead,  prescribing some 
probabilities to generated outcomes (i.e. considering the 
random sequences) we 
are able to compare the original and conjugate 
systems just through their analytical parameters (see 
Section 3 for details).  
The sequences $\nu_{k}$, being considered on some infinite
subsets of indices $\Lambda\subset N$ of the natural 
series $N$, are convergent to some compacts
from $(0,1)$, which can be disjoint for 
different $\Lambda$. For the sequences of independent 
random variables and Markov chains these cluster sets 
permit analytical description. 
Moreover, the analysis suggested can be applied to arbitrary
continuous-time stochastic systems, permitting approximation
or interpolation by the discrete ones. E.g., Section 3 gives 
such an application to diffusion stochastic processes.
As another example (does not considered in this work) one can
refer to Nyquist-Shannon sampling
theorem [8, 32] from data analysis, according to which the
analogue signal with bounded power spectrum is completely
determined through its values on some discrete set of equidistant
points.

\subsection{A new characteristic for irregular systems}

One of the main claims relating to the approach 
described, which has been confirmed by preliminary 
computations ([3]),  
is the following: a periodic response of given stochastic
system on a weak periodic perturbation is localized in
${\bar X}^{(n)}$ -- its presence
can be detected, when considering the higher order
differences, taken from initial orbit $X$.
For that purpose it should be studied the
asymptotical (as $N\to \infty$) relative
volume (the density in natural series) of the set of all
those indices $i$, for which the changes of binary symbol
occur,
\begin{equation}
  \delta_{i+1}^{(N)} = 1-\delta_{i}^{(N)}
\qquad \qquad (1\leq i\leq N-1).
\end{equation}
This leads to the following definition: For an orbit ${\bar X}=
(x_{k})_{k=0}^{\infty}$ of a given system we define
\begin{equation}
 \gamma = \gamma({\bar X}) = \lim_{N\to \infty}
 \frac{\gamma({\bar X}, N)}{N} \qquad \qquad (0\leq \gamma \leq
 1)
\end{equation}
where $\gamma ({\bar X}, N)$ denotes the total number of
those indices $1\leq i\leq N-1$ for each of which the 
(8) holds; the existence of limit in (9) is preassumed.

For deterministic systems the theoretical 
study of properties of $\gamma$ 
is quite difficult. 
Two statements of this section, Theorem 1
and Corollary 1, are apparently the only available
theoretical results. Some theoretical statements, relating
to random binary sequences can be found in Section 3.
On the other hand, the $\gamma$ has that important
advantage, that is the simplicity of its computation.
It can be easily implemented just over the (experimental)
data $\bar X$, without referring to the process generation
law. By this reason, as for
computation of $\gamma$ we need only
the corresponding time series to be available, this new
characteristic is well adapted for computational study
of various applied problems.

The numerical analysis shows (see [3]) 
that $\gamma$ has weak dependence on
system's initial values and is able to distinguish the 
regular and chaotic motion. 
With this aim we have compared $\gamma$ with Lyapunov
exponent $\lambda$ (see e.g., [12], Ch.~5 and
[4], Ch.~7.2.b): for a given map $F: (0,1) \ \to \ (0,1)$
it is defined as
\begin{equation}
\lambda = \lambda({\bar X})=\lim_{N\to \infty}
\frac{1}{N}\sum_{k=1}^{N}
\ln|\frac{dF(x_{k})}{dx_{k}}|
\qquad (x_{k+1}=F(x_{k})) \ .
\end{equation}
It is known [4] that Lyapunov exponent of any integrable
system is zero. In contrary, it follows from Theorem 1 below that
there exist integrable (but aperiodic) systems with positive
(and rational) $\gamma$ -- it can be, e.g., the sequence of
fractional parts. 
Some results on computations 
of $\gamma$-characteristic as well as its comparisons
with Lyapunov exponent can be found in [3].  
Three simple deterministic systems -- the tent
map, logistic function, and Poincare displacement
of Chirikov's standard map have been examined.  
The numerical results demonstrate a 
strong correlation of 
$\gamma$ with Lyapunov exponent (see [3], Figs.~1 and 2). 
 
We emphasize another important feature of this quantity, has been 
derived from these preliminary computations: 
the $\gamma$-coefficient is that numerical
characteristic, associated with a given stochastic system, which is
able to change significantly its numerical value when the
system undergoes a weak perturbation. 
This relates to stochastic
resonance phenomena. Some works [24, 25, 26] claim that
namely this resonance mechanism has the basic role in neural
activity. 

We have proven two rigorous results relating to 
$\gamma$-characteristic
of continuous deterministic systems -- Theorem 1 and Corollary 1.  
The Corollary 1 follows from Theorem 2, giving
some 'difference' analogy of Eggleston formula
from the ergodic theory.
If $\bar X$ is either constant or periodic, then we obviously
have
\begin{equation}
   \gamma({\bar X}, N)=\gamma N+O(1) \qquad (N\to \infty)
\end{equation}
and the coefficient $0\leq \gamma\leq 1$ is rational. 
According to next
theorem (the symbols $\{.\}$ and $[.]$ denote fractional and 
entire part of number) this remains valid also for sequences of 
fractional parts.
These (conditionally periodic) sequences have an important role on studying the general
integrable dynamical systems (see [34, 35]).
\newtheorem{guess1}{Theorem}
\begin{guess1}
For the sequence ${\bar X} = (\{\alpha n\})_{n=1}^{\infty}$
where $0<\alpha <1$ is irrational, the next statements are
true:
(1)~the conjugate to $\bar X$ orbit
$\bar\nu= (\nu_{n})_{n=1}^{\infty}$ is a periodic
sequence;
(2)~if entire part of $1/\alpha$ is of
the form $[1/\alpha]= 2^{p}-1$ ($p\geq 1$)
then $\nu_{n}\equiv 0$ for all large enough indices $n$.
\end{guess1}

The second theoretical result on $\gamma$-characteristic is the 
Corollary 1, establishing  for binary systems an upper estimate for Hausdorff 
dimension of the attractors $\cal A$ (see previous section 2.1). To formulate this 
estimate involving $\gamma$-characteristic and Shannon 
entropy function, we need some preliminary definitions and results. 
Let us consider the processes, generating the binary
sequences
\begin{equation}
   {\bar x} = (x_{1}, x_{2}, \ldots, x_{n}, \ldots),
   \qquad x_{i}\in \{0,1\} \ ;
\end{equation}
it is convenient to prescribe to such a sequence the number
$0<x<1$
$$ x = 0.x_{1}x_{2}\ldots x_{n}\ldots \qquad
(= \sum_{n=1}^{\infty}2^{-n}x_{n}) \ . $$
We define ${\cal B}_{K}$
as the collection of all real numbers $0<x<1$ for which the
sequences (12) contain only bounded (by a given number
$K$) series with
the same binary symbol. If all of the difference sequences
${\bar x}^{(k)}$ belong to ${\cal B}_{K}$, the $\bar x$
is called [2] $\beta_{K}$ sequence. 
Some necessary and sufficient conditions a sequence $\bar x$
to be a $\beta_{K}$ sequence can be found in [2].

The Eggleston theorem ([7]) states that
\begin{equation}
  dim\;(\{ x\in (0,1): \lim_{n\to\infty}
  \frac{1}{n}\sum_{i=1}^{n}x_{i} = p \}) = H(p)
\end{equation}
where notation $dim (E)$ stands for Hausdorff dimension of 
set $E$ and
\begin{equation}
  H(x)=x\log_{2}\frac{1}{x}+(1-x)\log_{2}\frac{1}{1-x} \qquad (0<x<1)
\end{equation}
is Shannon entropy function.
This function can also be derived 
from theory of number partitions: if $C(s,N)$
denotes the total number of
compositions of number $N$ into $s$ parts,
\begin{equation}
N = m_{1} + m_{2} + \cdots + m_{s} \ ,
\end{equation}
then since $C(s,N)=C_{N-1}^{s-1}$, using Sterling formula
for binomial coefficients, it can be easily obtained that
\begin{equation}
H(x)=\lim_{N\to\infty}\frac{1}{N}\log_{2} C(xN,N)
\qquad (0<x<1) \ .
\end{equation}
Analogously, given $K\geq 1$ we define
\begin{equation}
H_{K}(x)=\lim_{N\to\infty}\frac{1}{N}\log_{2} C_{K}(xN,N)
\qquad (0<x<1)
\end{equation}
where $C_{K}(s,N)$ denotes ([9]; [10], Ch.~4.2) the total
number of compositions (15) satisfying the 
restriction $m_{i}\leq K$.  
To be correct in these definitions,
one should counts $x$ is rational -- if $x=p/q$, then in (16)
and (17) $N$ is of the form $N=qn$ and $n\to \infty$.
Then the function (16) (as well as the (17)) can
be extended
on the whole unit interval $0<x<1$.
Concerning the relation (16) and the  
procedure just used
for defining the Shannon entropy, see also [36, 37].

The next Theorem 2 estimates the Hausdorff dimension of the
sets
\begin{equation}
  E(p)=\{ x\in (0,1): \lim_{n\to\infty}
  \frac{1}{n}\sum_{i=1}^{n}|x_{i+1}-x_{i}| = p \}
\end{equation}
\begin{equation}
  E_{K}(p)=\{ x\in {\cal B}_{K}: \lim_{n\to\infty}
  \frac{1}{n}\sum_{i=1}^{n}|x_{i+1}-x_{i}| = p \}
\end{equation}
where $0\leq p\leq 1$ and $K\geq 1$ are arbitrary.
\begin{guess1}
The next inequalities
\begin{equation}
   dim\;(E(p)) \leq H(p),  \qquad
   dim\;(E_{K}(p)) \leq H_{K}(p)
\end{equation}
are true.
\end{guess1}
The following result establishes some relationship between system's $\gamma$-characteristic and Hausdorff dimension of its conjugate attractor.
\newtheorem{guess2}{Corollary}
\begin{guess2}
If a deterministic system generates the binary sequences
(the binary $\beta_{K}$ sequences), then for Hausdorff
dimension of the attractor $\cal A$ we have
\begin{equation}
dim ({\cal A})\leq H(\gamma) \qquad
(dim ({\cal A}) \leq H_{K}(\gamma))
\end{equation}
where $\gamma$ is the system's response characteristic
defined by Eq.~(9).
\end{guess2}

The $H$ in (21) is Shannon entropy function, having a simple 
analytic form (14). We do not have such a simple expression for 
the function $H_{K}$. 
The generating function of numbers $C_{K}(s,N)$ from (17), 
by means of which the $H_{K}$ is defined, 
is well known ([10], Ch.~4.3):
$$ \sum_{N=0}^{\infty}C_{K}(s,N)q^{N} =
(q+q^{2}+\cdots + q^{K})^{s} \ ;  $$
using this formula, as in [10], Ch.~4.3, it can be obtained
 $$C_{K}(s,N) =
 \sum_{p=0}^{s}(-1)^{p}C^{p}_{s}C^{s-1}_{N-K-1} \ . $$
Some asymptotic relations for this quantity can be found in [11]
(see [10], Ch.~4, Comments; the work [11] remained 
unavailable to us). They can be used to derive the
explicit analytic form of the function $H_{K}(x)$.
In respect of inequality (20) we note also the
work [29], where the positive discrepancy of fractal measures
from Shannon entropy is discussed.

\subsection{Conjugate orbits and bifurcation points}

The conjugate orbits in fact magnify the small fluctuations 
of the process.
The fluctuations are usually occur at unstable branch
points [5, 6]. The notion of conjugate orbit allows 
to determine 
that each branch point on the bifurcation diagram
of the original system can be treated as
the source of chaos for some 'shifted' conjugate 
system.

The computations were made for logistic map $T: x \to rx(1-x)$.  
Namely, let
$$ b_{1}< b_{2}< \cdots <b_{k} < \cdots $$
be the bifurcation points
of the orbit $X = (x_{n})_{n=1}^{\infty}$ of iterates 
$x_{n+1} = T(x_{n})$  $(n\geq 1)$,
numbered in increasing order. It is known [15], that
  $$\lim_{k\to \infty} b_{k} = b_{\infty}
  = 3.5699\ldots  \ \ . $$
Given $N_{k}$ tending to infinity we consider the shifted
sequences ${\bar Y}_{k} = (x_{n+N_{k}})_{n=1}^{\infty}$.
The claim is that if $N_{k}$ goes to $\infty$ quickly
enough, then the orbits ${\bar\nu}_{k}$, conjugate to
${\bar Y}_{k}$, demonstrate the chaotic behavior at points
$b_{1}, b_{2}, \ldots , b_{k}$, while they are
identically zero for all the different values of control parameter 
$r$ belonging to interval $(0, b_{\infty})$.

This is certain algorithmic formalism 
of the mentioned in Section 1 descriptive remarks
from [5] and [6] relating to small fluctuations. 
The same analysis can be
implemented for many other systems exhibiting the bifurcations.
Thus, the bifurcation diagrams of Poincare displacement
of Duffing equation ([13], Ch.~11.5), Rossler system
([41], p.46),
and forced magnetic oscillator ([12], Ch.~2) are
very similar to that of the considered logistic map
([14], Ch.~3).

The bifurcations are often treated as the chaos precursors.
Hence, any methods for their detection are of a great
practical interest (see, e.g. [12]).

\section{Random processes}

\subsection{Sequences of random independent variables}

For the case of random processes the difference analysis has 
richer consequences and permits theoretical study. We consider
discrete random processes of the form
\begin{equation}
\bar \xi = (\xi_{1}, \xi_{2}, \ldots, \xi_{n}, \ldots)
\end{equation}
coordinates $\xi_{n}$ of which take binary values
$0$ and $1$ with some positive probabilities,
$$ P(\xi_{n}=0) = p_{n}, \qquad P(\xi_{n}=1) = q_{n} \qquad
(p_{n}+q_{n}=1) \ . $$
Then the differences $\xi_{n}^{(k)}$,
$$
\xi_{n}^{(k)} = |\xi_{n+1}^{(k-1)} - \xi_{n}^{(k-1)}|
\qquad (n,k\geq 1, \ \bar\xi_{n}^{(0)}=\xi_{n})
$$
also take binary values with some positive probabilities
$$ P(\xi_{n}^{(k)}=0) = p_{n}^{(k)}, \qquad
   P(\xi_{n}^{(k)}=1) = q_{n}^{(k)}  \qquad
(p_{n}^{(k)}+q_{n}^{(k)}=1) \ . $$
Hence, one can consider the random difference processes
\begin{equation}
{\bar \xi}^{(k)} =
(\xi_{1}^{(k)}, \xi_{2}^{(k)}, \ldots, \xi_{n}^{(k)},
\ldots ) \ ;
\end{equation}
we will also deal with the corresponding random variables
of the form 
$${\bar \xi}^{(k)} =
0.\xi_{1}^{(k)}\xi_{2}^{(k)}\ldots\xi_{n}^{(k)}\ldots  \qquad
(k\geq 0, \  \xi_{n}^{(0)}\equiv \xi_{n}, \ \xi^{(0)}\equiv \xi) $$
(the notation is the same as for sequences (23)).
It is easy to see that
 $${\bar\xi}^{(k)} = {\cal R}^{(k)}\;{\bar \xi}$$
where ${\cal R}^{(k)}$ is $k$-th iterate of the
"fractal map" $\cal R$ from [2]:
$$ {\cal R}\;{\bar \xi} = 0.\;{\xi_{1}}\oplus{\xi_{2}}\;
{\xi_{2}}\oplus{\xi_{3}}\ldots
{\xi_{n}}\oplus{\xi_{n+1}} \ldots $$
(we use notation
$\alpha \oplus \beta\  (=|\alpha - \beta|)$ for logical
sum of binary variables $\alpha$ and $\beta$).

We are interested in the limiting behavior of these differences
when $k$ goes to infinity. We
say that ${\bar\xi}^{(k)}$ converges to a random sequence
${\bar\xi}^{(\infty)}$ if the $p_{n}^{(k)}$ tend
to some numbers
$p_{n}^{(\infty)}$ as $k\to \infty$ and
$k\in \Lambda$ (convergence by probability);
here $\Lambda$ is a given infinite subset of natural
series and the final probabilities may depend on $\Lambda$,
$p_{n}^{(\infty)} = p_{n}^{(\infty)}(\Lambda)$.
Then
$${\bar\xi}^{\infty} = {\bar\xi}^{(\infty)}_{\Lambda} =
(\xi_{1}^{(\infty)}, \xi_{2}^{(\infty)}, \ldots,
\xi_{n}^{(\infty)}),   \qquad \xi_{n}^{(\infty)} =
\xi_{n}^{(\infty)}(\Lambda) $$
is again a discrete random process, binary components
of which take 
values $0$ and $1$ with some final (or stationary --
if follow the terminology of Markov processes) probabilities,
$$ P(\xi_{n}^{(\infty)}=0) = p_{n}^{(\infty)}, \qquad
   P(\xi_{n}^{(\infty)}=1) = q_{n}^{(\infty)}  \qquad
(p_{n}^{(\infty)}+q_{n}^{(\infty)}=1) \ . $$

The following 3 statements are the basic
tools we apply for studying the limiting
behavior of differences of binary random sequences. For
$\epsilon_{i}\in \{0,1\}$ we use the notation
 $$ <\epsilon_{0}, \epsilon_{1}, \ldots, \epsilon_{k}> =
    (\sum_{i=0}^{k}\epsilon_{i}C^{i}_{k})\;mod\;(2) \ . $$
\newtheorem{guess}{Lemma}
\begin{guess}
For probabilities of $k$-th ($k\geq 0$) differences
we have
\begin{equation}
 P(\xi_{n}^{(k)} = \lambda) =
 \sum_{<\epsilon_{0},\epsilon_{1}\ldots \epsilon_{k}> =\lambda}
 P(\xi_{n}=\epsilon_{0})P(\xi_{n+1}=\epsilon_{1})
 \cdots P(\xi_{n+k}=\epsilon_{k})
\end{equation}
where $n\geq 1$ and $\lambda \in \{0,1\}$ are arbitrary.
\end{guess}

It is convenient to represent the probabilities
in the form
 $$
 P(\xi_{n} = \lambda) =
 \frac{1}{2}(1+(-1)^{\lambda}\pi_{n}), \qquad
 P(\xi_{n}^{(k)} = \lambda) =
\frac{1}{2}(1-(-1)^{\lambda}\pi_{n}^{(k)})
 $$
where $-1\leq \pi_{n}, \pi_{n}^{(k)} \leq 1$
are some numbers.
The proof of Lemma 2 is based on the following remark:
\newtheorem{guess3}{Remark}
\begin{guess3}
The identity
$$
\sum_{<\epsilon_{0},\epsilon_{1},\ldots,\epsilon_{k}>=\lambda}
x_{0}^{\epsilon_{0}}x_{1}^{\epsilon_{1}}\cdots
x_{k}^{\epsilon_{k}} = \frac{1}{2}[1+(-1)^{\lambda}
\prod_{0\leq i\leq k, \  \alpha_{i,k} = 1}^{k}
\frac{1-x_{i}}{1+x_{i}}]\prod_{i=0}^{k}(1+x_{i})
$$
is true.
\end{guess3}

The following lemma gives an explicit
expression of $\pi_{n}^{(k)}$ by means of $\pi_{n}$.
To formulate it, we consider the binary analogy
$P = (\alpha_{i,k})_{i=0,n; \ \ k=0,\infty}$ of Pascal
triangle of binomial coefficients, is defined as follows:
$\alpha_{0,k} = \alpha_{k,k}=1$ and
  $$  \alpha_{i,k} = \left \{   \begin{array}{ll}
      0, & C_{k}^{i} \quad \mbox{is even}\\
      1, & C_{k}^{i} \quad \mbox{is odd}
      \end{array} \right.
      \quad \mbox{i.e.} \quad \alpha_{i,k}
      = \alpha_{i-1,k-1}\oplus \alpha_{i,k-1} \ .
  $$
for $1\leq i\leq k-1$.
The fractal graphical image of the triangle $P$
can be found in [15], where it is considered in
connection of cellular automata theory.
\begin{guess}
For arbitrary $n, k\geq 1$ the equality
\begin{equation}
\pi_{n}^{(k)} = \prod_{0\leq i \leq k,
\ \alpha_{i,k}=1}\pi_{n+i}
\end{equation}
is true.
\end{guess}

We are interested in the existence of limit of the
$\pi_{n}^{(k)}$ when $k\to \infty$. When $k$ converges to
infinity arbitrarily , this limit may not exist. For
instance, in the simplest case $p_{n}\equiv p$
(or $\pi_{n}\equiv \pi$), it follows from Remark 2 below,
that if the binary
code of number $k$ contains exactly $m$ units, then
$\pi_{n}^{(k)} = \pi^{2^{m}}$. Since for every given
$s$ the collection $\Lambda_{m}$
of all such numbers $k$ is infinite, it is clear that the
limit mentioned, generally speaking, does not exist.
On the other hand, arbitrary given
$\pi_{n}$ the Lemma 2 in principle
allows to describe all of the infinite subsets
$\Lambda \subset N$ for which the limit
  $$\pi_{n}^{(\infty)}= \pi_{n}^{(\infty)}(\Lambda) =
  \lim_{k\to \infty, \ k\in\Lambda} \pi_{n}^{(k)}$$
exists. In other words, Lemma 2 provides
a sufficient tool
to describe all of the $\Lambda$ for which the limiting
random sequence $\xi^{(\infty)}(\Lambda)$ exists:
it follows from (25) that for any infinite $\Lambda$
and $n\geq 1$ the final
probabilities can be computed by formula
\begin{equation}
    \ln \frac{1}{|\pi_{n}^{(\infty)}(\Lambda)|} =
    \lim_{k\to \infty, \  k\in \Lambda} \sum_{0\leq i\leq k,
    \  \alpha_{i,k}=1} \ln \frac{1}{|\pi_{n+i}|}
\end{equation}
provided the right hand limit exists (or is infinite).

The most of further results relate to
studying these final processes for some particular case:
we consider the limiting transition of the differences 
${\bar\xi}^{(k)}$ as $k\to \infty$ and $k\in \Lambda_{0}$
where
 $$ \Lambda_{0} = \{k=2^{p}-1: p=1,2,\ldots \} \ . $$
Then for arbitrary $k$ we have all of the 
$\alpha_{i,k}=1$ and hence the relation (25) gains a
simpler form
\begin{equation} 
 \pi_{n}^{(k)} = \prod_{i=0}^{k}\pi_{n+i} \ . 
\end{equation}
From where immediately follows:
\begin{guess1}
The limiting process ${\bar\xi}^{(\infty)}(\Lambda_{0})$
is the
symmetric random walk, i.e. $\pi_{n}^{(\infty)} \equiv 0$, 
if and only if when
\begin{equation}
\sum_{n=1}^{\infty}\ln \frac{1}{|\pi_{n}|} = \infty \ .
\end{equation}
In the contrary case, when this series is convergent, we
have
\begin{equation}
|\pi_{n}^{(\infty)}||\pi_{1}^{(\infty)}|^{-1} =
\prod_{i=1}^{n-1}|\pi_{i}|^{-1} \qquad
(=|\pi_{n}||\pi_{n+1}||\pi_{n+2}|\cdots ) \ .
\end{equation}
\end{guess1}

It is clear that if $p_{n}\equiv const$, then (28) holds.
If $\pi_{n} = 2^{-n}$ $(n\geq 1$), we have an example 
of a self-conjugate system:
${\bar\xi}^{(\infty)}\equiv \bar\xi$.
For Poisson distribution, when
$p_{n}=e^{-\lambda}\lambda^{n}/n!$, using (27) it can be
easily computed 
$$|p_{n}^{(\infty)} - p_{n}| = o\;(p_{n})
\qquad (n\to \infty) \ . $$
The same relation is valid also for Poisson
homogeneous events flow with probabilities
$P_{n}(t)=e^{-\lambda t}(\lambda t)^{n}/n!$.
Indeed, (following [16], Ch.~6.5) for a given $t$ we
divide time interval $[0,t]$ into equal intervals of
lengths $1/n$ and on the obtained finite lattice
consider Bernoulli's trial scheme with success
probability is equal to $\lambda t/n$.
Since the success probability in $s$ trials tends
(as $n\to \infty$) to the function $P_{n}(t)$, we
obtain 
\begin{equation}
  |P_{n}^{(\infty)}(t) - P_{n}(t)| =
  o\;(P_{n}(t))  \qquad (n\to \infty) \ .
\end{equation}

The next statement is an immediate consequence of Theorem 3.
\begin{guess2}
The random sequence
$\bar\eta = (\eta_{1}, \eta_{2}, \eta_{3}, \ldots )$
is a limiting (for some $\bar\xi$, i.e.
$\bar\eta = {\bar\xi}_{\Lambda_{0}}^{(\infty)}$)
process, if and only if when either $\bar\eta$ is
the symmetric random walk or the sequence
$|\pi_{n}(\bar\eta)|$ monotone increases to $1$ as
$n\to \infty$.
\end{guess2}
\begin{guess2}
If $\bar\eta$ is a limiting random sequence (for some
$\bar\xi$, i.e.
$\bar\eta = {\bar\xi}_{\Lambda_{0}}^{(\infty)}$), differing
from symmetric random walk, then the random variable
 $$ 0.\eta_{1}\eta_{2}\eta_{3}\ldots $$
has a pure singular probability distribution.
\end{guess2}

The latter statement follows from
Marsaglia's results [17] -- 
Marsaglia's criterion a random variable
$\bar\eta$ possesses a singular distribution, is
$$\sum_{n=0}^{\infty}|\pi_{n}(\bar\eta)|^{2} =
\infty   \ . $$
The Corollary 2 provides a stronger condition:
$|\pi_{n}(\bar\eta)| \to 1$.
Since we have
$$ 1-|\pi_{n}| = 2\min(p_{n}, q_{n}) \ , $$
the convergence of series in (28) is equivalent to condition
 $$\sum_{n=1}^{\infty}\min\;(p_{n}, q_{n}) < \infty $$
According to [17], this is also equivalent to a requirement the
random variable $\bar\eta$ from Corollary 3 has a discrete 
probability density function.

If $\pi_{n}\equiv const$, then according to Theorem 3
the ${\bar\xi}^{(\infty)}_{\Lambda_{0}}$ is the symmetric
random walk. It
can be shown that for this case Theorem 3 remains
valid when the limiting transition is taken over some
"wide" subsets $\Lambda$ of natural series:
\begin{guess1}
If $\pi_{n}\equiv const$, then there exists a set
$\Lambda \subset N$
with density $1$ in natural series, such that
${\bar\xi}^{(\infty)}_{\Lambda}$ is the symmetric
random walk.
\end{guess1}

Here, as such a $\Lambda$ it can be chosen a sequence of
natural numbers, for which the total number of units in
their binary codes increases to infinity quickly enough.
Theorem 4 follows from the next proposition:
\begin{guess3}
(1)~The total number of units in $k$-th line of binary
Pascal triangle is equal to $2^{m(k)}$ where $m(k)$ is
the total number of units in binary code of number $k$.
(2)~There exist sets $\Lambda \subset N$ such that
$\lim_{k\to \infty, \ k\in \Lambda}m(k)=+\infty$
and $dens (\Lambda)=1 $.
\end{guess3}

For a given $\bar\xi$ we let
\begin{equation}
  \gamma({\bar\xi},k) = \xi_{1}^{(k)} + \xi_{2}^{(k)} +
  \cdots + \xi_{k-1}^{(k)}
\end{equation}
-- the total number of units in (the realizations of)
$k$-th difference sequence
$(\xi_{1}^{(k)}, \ldots, \xi_{k-1}^{(k)})$;
it also coincides with
the total number of changes of the binary symbol in the sequence
$(\xi_{1}^{(k-1)}, \ldots, \xi_{k}^{(k-1)})$:
\begin{equation}
   \gamma({\bar\xi},k) = |\xi_{1}^{(k-1)} - \xi_{2}^{(k-1)}| +
  \cdots + |\xi_{k-1}^{(k-1)}-\xi_{k}^{(k-1)}| \ .
\end{equation}
The next two statements follow from Theorems 3, 4 and Remark 2 
and relate to $\gamma$-characteristic of random sequences:
as in (9), given infinite $\Lambda\subset N$, we define    
\begin{equation}
 \gamma({\bar\xi},\Lambda) = \lim_{k\to \infty, \  k\in \Lambda}
 \frac{\gamma({\bar\xi}, k)}{k}
\end{equation}
where the existence of the limit is assumed.  
We note again that the
limit in (33) with $\Lambda \equiv N$ , generally speaking 
may not exist and then one has to study the corresponding 
non-trivial cluster sets of the ratio in (33). 
The same situation holds also for Lyapunov exponent, 
is defined by Eq.~(10).   

We consider some particular case of processes $\bar\xi$ from  
(22),  requiring 
\begin{equation}
 P(\xi_{n}=0)\geq P(\xi_{n}=1)
\end{equation}
or what is the same,  $\pi_{n}\geq 0$ for all $n\geq 0$.
\begin{guess1}
Let us have a random process $\bar\xi = (\xi_{n})_{n=1}^{\infty}$ satisfying (34). If
$$
\sum_{n=0}^{\infty}P(\xi_{n}=1) <\infty \quad 
\mbox{then with probability 1} \quad
  \lim_{k\to \infty, \ k\in \Lambda_{0}}
  \frac{\gamma ({\bar\xi}, k)}{k} = 0 $$
and if
$$\sum_{n=0}^{\infty}P(\xi_{n}=1) = \infty \quad
\mbox{then with probability 1} \quad
\lim_{k\to \infty,\ k\in \Lambda_{0}}
\frac{\gamma ({\bar\xi}, k)}{k} = \frac{1}{2}  \ . $$
\end{guess1}
\begin{guess2}
If $\pi_{n}\equiv \pi$ then
there exists $\Lambda\subset N$ such that
$dens (\Lambda)=1$ and $\gamma ({\bar\xi},\Lambda)= 1/2$.
\end{guess2}

\subsection{Conjugate attractors of identically
distributed sequences}

On considering the random sequences (22), one
may assume ([16], Ch.~8.6 and  [22], Ch.~4.3) that
\begin{equation}
 \xi_{n}=\xi_{n}(\omega) =
 \omega_{n} \quad \mbox{where} \quad
 \omega=0.\omega_{1}\omega_{2}\omega_{3}\ldots  \ \ ;
\end{equation}
here $\omega$ is real number from unit interval, given in
form of
its binary expansion. We are interested in Hausdorff
dimension of the sets
\begin{equation}
  M = M(p) = \{ \omega\in
  [0,1]:\lim_{n\to\infty}\frac{1}{n}
  \sum_{k=1}^{n}\xi_{k}(\omega) = p \}
\end{equation}
where is assumed $P(\xi_{n}=0)=p=const$.
If $p=1/2$, then $M$ is
the set of so-called Borel normal numbers ([7])
for which $mes(M)=1$ and
$dim (M)$ =1 ($mes$ is Lebesgue measure on $[0,1]$).
Hence, if in (36) $p\neq 1/2$, then $mes(M)=0$. However,
in that case there exists a singular
measure $\nu$ on interval $[0,1]$ such that $supp (\nu) = M$,
$\nu (M) = 1$ ([7], Ch.~4). The Eggleston theorem
provides us with explicit expression for Hausdorff dimension:
\begin{equation}   
           dim (M(p)) = H(p) 
\end{equation} 
where $H$ is Shannon function (14). 
The derivative of distribution function of measure
$\nu$, i.e. the probability density of the random variable
 $$0.\xi_{1}(\omega)\xi_{2}(\omega)\xi_{3}(\omega)
 \ldots $$
is a singular function, concentrated on a set of zero
Lebesgue measure. The self-affine graph of such a function
can be found in [7], Ch~3.1.

For a given stochastic system it is interesting to describe all
of the attractors ${\cal A}_{m}$ (we assume they are 
numbered by increase of their Hausdorff dimension), mentioned 
in section 2.1. It can be easily done for the case
$\pi_{n}\equiv const \ (= \pi)$. Indeed, if we let
\begin{equation}
  \Lambda_{m} =
  \{ k \in N: k=2^{p}+m, \  p=0,1,2,\ldots \}
\end{equation}
then according to Remark 2 for the limiting
process ${\bar\xi}_{\Lambda_{m}}^{(\infty)}$ we 
have $\pi_{n}^{(\infty)}(\Lambda_{m}) \equiv \pi^{2^{m}}$.
For the terms of conjugate orbit $\bar\nu$ we have
 $$
 \nu_{k}=0.\xi_{1}^{(k)}\xi_{2}^{(k)}\xi_{3}^{(k)}\ldots
 \ .
 $$
From where it is clear that the conjugate attractor
${\cal A}_{m}$
corresponding to random variable ${\bar \xi}^{(\infty)} =
{\bar\xi}^{(\infty)}_{\Lambda_{m}}$, coincides with
the set, on which the sequence
  $$ {\bar\xi}^{(\infty)}_{\Lambda_{m}} =
0.\xi_{1}^{(\infty)}\xi_{2}^{(\infty)}\xi_{3}^{(\infty)}
\ldots
  $$
is localized (with probability $1$).
Then Eggleston theorem implies  
\begin{guess1}
For the sequence of
identically distributed random variables, $\pi_{n}\equiv \pi$,
the set of all its conjugate attractors 
is a collection of some Eggleston sets, 
\begin{equation}
  {\cal A}_{m} = M(\frac{1+\pi^{2^{m}}}{2}) \qquad 
(m=0, 1, 2, \ldots) \ .
\end{equation} 
\end{guess1}

If we let   
 $$ {\hat H}(x) =
    H(\frac{1+x}{2}) \qquad ({\hat H}(x) = {\hat H(-x)}, 
\quad 0\leq {\hat H}(x)\leq 1), \quad -1\leq x\leq 1  $$
then Theorem 6 gives 
 $ dim({\cal A}_{m}) = {\hat H}(\pi^{2^{m}}) $, 
hence 
${\hat H}^{-1}(dim ({\cal A}_{m}))=\pi^{2^{m}}$ and    
thus the next statement is true:
\begin{guess1}
For Hausdorff dimensions of conjugate attractors ${\cal A}_{m}$ 
of identically distributed random sequence
the equalities
\begin{equation}
({\hat H}^{-1}(dim ({\cal A}_{1}))^{1/2} =
({\hat H}^{-1}(dim ({\cal A}_{2}))^{1/4} = \cdots =
({\hat H}^{-1}(dim ({\cal A}_{m}))^{1/2^{m}} = \cdots
\end{equation}
hold.
\end{guess1}

\subsection{Binary Markov processes} 

Here we give analogies of some results from Sec.~3.1       
for the case of infinite binary Markov chains.  
\begin{guess3} 
Let  ${\bar\xi}= (\xi_{n})_{n=1}^{\infty}$ be 
a binary Markov chain with the transaction 
probabilities $\pi_{n}(x,y) = P(\xi_{n}=y|\xi_{n-1}=x)$ 
and with the probabilities $p_{n}(x,y) = P(\xi_{n}=x)$ of 
attainment the value $x$ for $n$ steps by the finite chain 
$(\xi_{k})_{k=1}^{n}$. Then for every $k\geq 1$ the 
difference sequence
$\bar\xi^{(k)} = ({\bar \xi}_{n}^{(k)})_{n=1}^{\infty}$ 
is also a Markov chain and for the corresponding 
probabilities $\pi_{n}^{(k)}(x,y)$ and $p_{n}^{(k)}(x)$ we 
have the following recurrent relationships: 
\begin{equation} 
p_{n}^{(k)}(x) = p_{n-1}^{(k)}(0)\pi_{n}^{(k)}(0,x) + 
p_{n-1}^{(k)}(1)\pi_{n}^{(k)}(1,x) 
\end{equation} 
\begin{equation} 
\pi_{n}^{(k)}(x,y)=\pi_{n-1}^{(k-1)}(0,x)\pi_{n}^{(k-1)}(x,|x-y|) 
+  \pi_{n-1}^{(k-1)}(1,1-x)\pi_{n}^{(k-1)}(1-x,1-|x-y|) 
\end{equation} 
\end{guess3}

For the general Markov processes, the computation of  
distributions $\pi^{(\infty)}$ and $p^{(\infty)}$ for 
the limiting processes 
${\bar\xi}^{(\infty)}_{\Lambda}$ can be complicated. 
However, we are able to compute the distribution $p^{(\infty)}$ for 
homogenous processes and for arbitrary given 
$\Lambda = \Lambda_{s} = \{2^{p}+s: p=0, 1, 2, \ldots \}$. 
This is  based on the next two statements. 
\begin{guess} 
If $\bar\xi = (\xi_{n})_{n=1}^{\infty}$ is a homogenous 
Markov process and $\Lambda=\{2^{p}: p\geq 0\}$ then for 
the difference processes 
${\bar\xi}^{(k)} = (\xi_{n}^{(k)})_{n=1}^{\infty}$, $k\in \Lambda$,  
we have   
\begin{equation} 
P(\xi_{m}^{(k)}=\lambda) = q^{k}\sum_{\epsilon, \delta \in\{0,1\}}
((\frac{q}{1-q})^{\epsilon+\delta-1}\sum_{[p/2]=\lambda}
\sum_{k=1}^{\infty}C_{m-p}^{k-\epsilon -\delta}C_{p-1}^{k-1}x^{p}y^{k} 
\end{equation} 
where $\lambda\in \{0,1\}$, $x=\frac{s}{q}$, 
$y=\frac{1-s}{s}\frac{1-q}{q}$, $s=p(1,1)$, $q=p(0,0)$ and 
$p(x,y)$ is the transaction 
probability function for the process $\bar\xi$.  
\end{guess} 
\begin{guess3}
The next identity 
\begin{equation} 
\sum_{p=0}^{\infty}\sum_{k=0}^{\infty}C_{p}^{k}
C_{m-p}^{k}x^{p}y^{k} = 
(1+x)^{m}T_{m}(\frac{x(y-1)}{(1+x)^{2}}) 
\end{equation} 
where $T_{m}(z)= \sum_{k=0}^{m}C_{m-k}^{k}z^{k}$ is Chebyshev polynomial, 
is true. 
\end{guess3} 
These three statements imply:
\begin{guess1} 
If ${\bar\xi} = (\xi_{n})_{n=1}^{\infty}$ is a homogenous 
Markov process and $\Lambda = \{ 2^{p}: p\geq 0\}$, then we have 
\begin{equation} 
\lim_{k\in \Lambda, \ k\to \infty}P(\xi_{n}^{(k)}= \lambda) = 
|\lambda-\frac{2(1-s)(1-q)}{(1-s)+(1-q)}| 
\end{equation} 
where $\lambda$, $s$, and $q$ are the same as in Lemma 3. 
\end{guess1} 
It is clear from (42) that if $\bar\xi$ is homogenous process 
then for every $k\geq 1$ the difference process ${\bar\xi}^{k}$ 
is again homogenous. It is clear now that the limits of the 
quantities $P(\xi_{n}^{(k)} = \lambda)$ can be computed for 
all the sets of indices of the form 
$\Lambda_{s}= 2^{p}+s: p=0,1,2 \ldots \}$. Indeed, it is not 
difficult to see from Theorem 8, that this limit is 
\begin{equation} 
\lim_{k\in \Lambda_{s}, \ k\to \infty}P(\xi_{n}^{(k)} = \lambda) = 
|\lambda - \frac{2\pi_{0}^{(s)}(1,0)\pi_{0}^{(s)}(0,1)} 
{\pi_{0}^{(s)}(1,0) +\pi_{0}^{(s)}(0,1)}| 
\end{equation} 
It is also clear that the quantities $\pi_{0}^{(s)}$ are 
easily determined recurrently according to (41) and (42) 
(and hence their explicit computation is also possible).  
Thus we have the convergence of the quantities 
$P(\xi_{n}^{(k)} = \lambda)$ along the $\Lambda_{s}$, however 
we cannot confirm the convergence for the probabilities 
$\pi_{n}^{(k)}(x,y)$. If their convergence is known apriori, 
the relation (41) implies 
\begin{equation} 
\frac{\pi_{n}^{(\infty)}(0,1)}{\pi_{n}^{(\infty)}(1,0)}= 
|1-\frac{1}{2}\frac{\pi_{n}^{(0)}(0,1) + \pi_{n}^{(0)}(1,0)}
{\pi_{n}^{(0)}(1,0)\pi_{n}^{(0)}(0,1)}| 
\end{equation}       
-- this ratio can measure the 'strength' of Markov dependence 
property for the final process.

\subsection{One-dimensional diffusion}

On obtaining the relations (30) for Poisson events flow, 
we used
the circumstance that this process can be approximated
by some
discrete random processes. By the same way, one can
obtain such a 
result for the diffusion processes. However, in this case
we need to consider some "averaged" differences
${\hat\xi}_{n}^{(k)}$, are defined as
$$P({\hat\xi}_{n}^{(k)} = 0) =
\frac{1}{2}(1-{\hat\pi}_{n}^{(k)})
\quad \mbox{where} \quad
{\hat\pi}_{n}^{(k)}=(\pi_{n}^{(k)})^{1/k} =
(\pi_{n}\pi_{n+1}\ldots\pi_{n+k})^{1/k} \ . $$
We will also need to use the $(C,1)$ summation by
Cesaro [18]:
 $$ (C,1) \int_{x}^{+ \infty}fds = \lim_{z\to
 + \infty}\frac{1}{z}\int_{x}^{z}fds  \ . $$
Indeed, let the diffusion process $\bar\xi(t)$ is
determined
by Fokker-Plank-Kolmogorov equation ([16, 19, 20])
\begin{equation}
\frac{\partial f(x,t)}{\partial t} =
-b \frac{\partial f(x,t)}
{\partial x} +\frac{a}{2}\frac{\partial^{2}
f(x,t)}{\partial x^{2}}
\end{equation}
where $a=a(x)$ and $b=b(x)$ are some coefficients
and $f(x,t)$ is
the probability density of diffusing particle. It is
well known (see e.g., [16], Ch.~5.4 and [19]) that such
processes can be obtained as a result of some limiting
transition from random walk on one-dimensional lattice:
we consider the lattice $L_{h}$ with a step $h>0$
and assume that a particle
changes its position on the discrete time moments are
proportional to some small $\tau>0$ and with the probabilities
\begin{equation}
p=\frac{1}{2}(1-\pi), \qquad q=\frac{1}{2}(1+\pi) \quad
 \mbox{where} \quad
  \pi=\frac{1}{2}\frac{b}{a}h  \ .
\end{equation}
Imposing the restriction $h^{2}/\tau \to a$ (it can be
assumed $a\equiv 1$) and based on
Central Limit Theorem, one can deduce the FPK-equation
as well as to obtain its solution.

The scheme we apply to study the difference structure
of such continuous processes is the following.
If we have a discrete motion ${\bar\xi}_{h}$ on the 
lattice $L_{h}$, the 
transition probabilities of which permit an equality of
the form
\begin{equation}
\pi_{n}=C_{n}h
\end{equation}
then, after calculations by formula (27), we transform
the quantities $\pi_{n}^{(k)}$ for
probabilities of $k$-th difference process taken from
random walk ${\bar\xi}_{h}$ to a form
$$ {\hat \pi}_{n}^{(k)} =
C_{n,k}h \quad (= C_{x,k}h) $$
where the coefficients $C_{x,k}$ are such that there exists
the 
$\lim_{k\to\infty}C_{x,k} = C_{x}$. In such a way
we can consider a random walk, corresponding to the
discretized difference process, assigned on
the same lattice $L_{h}$ and with the same time scale
$\tau$. For the considering case (49) we have
${\hat\pi}_{n}^{(k)}=\pi_{n}=bh/2$, i.e.
${\hat\xi}_{n}^{(k)}\equiv {\bar\xi}_{h}$. From where we
conclude that ${\hat\xi}^{(\infty)}(t)$ coincides with
$\bar\xi(t)$. One can  see,  the same
arguments lead to the following formula:
\begin{equation}
|B(x)| =
\exp (- \  (C,1)\int_{x}^{\infty}\ln |b(z)|dz) \ .
\end{equation}
By such a way, the next statement is true:
\begin{guess1}
The limiting averaged differential process
$\hat\xi^{(\infty)}(t)$,
taken from  the diffusion $\bar\xi(t)$ is again a 
diffusion
process. If $\bar\xi$ satisfies equation (48) then
the drift
coefficient $B$ for ${\hat\xi}^{\infty}$ can be computed
by (51).
\end{guess1}

The same analysis is also applicable to general birth
and death processes [16] as well as to abstract
diffusion processes, considering in non-standard
stochastic analysis [30] including the Ising model.
We note another possible applications of the above
given approach. It is queuing analysis in 
computer networks (see e.g., [20, 21]). This theory
deals
with randomly arriving demands to some processors. The
random
process of time intervals between arrival moments is
studied.
On investigating of so-called heavy traffic [20, 21], a great
importance has the diffusion approximation  -- the
process
of queue lengths can be approximated by diffusion process
(48) (cp. Wiener's neuron from next section).
This area of applications is as
large, that requires an independent study.

\section{Neural networks}

\subsection{Stochastic neuron models}

Wiener's model of neuron ([8], p.299) has been suggested
by Mandelbrot and Gerstein (see [23, 40]). On the
studying the actual neuron spike trains, they found that
for some
instances these trains are well approximated by (bounded)
diffusion
process. The idea is that the excitatory and inhibitory
signals,
arriving to neuron's input, can be mathematically
interpreted as a (bounded) random walk on one-dimensional
lattice with a small constant step $h$. The limiting
(as $h\to 0$) process $\bar\xi$ is described
by Ito's stochastic differential equation ([19], Ch.~5.4)
\begin{equation}
 d{\bar\xi}(t) = \mu(x,t)dt + \sigma(x,t)d{\bar W}(t)
\end{equation}
where $\bar W$ is Wiener's process.
These authors have found that for some appropriately chosen
values of parameters $\mu$ and $\sigma$, the process
$\bar\xi(t)$  fits well the experimental data on
neural activity. The equation (52) is equivalent to
diffusion equation (48) with $\mu\equiv b$ and
$\sigma^{2}\equiv a$ ([8], p.299).

It is noted ([8], p.299) that inclusion a noisy
component to a given deterministic system may cause
the diffusion process in the system.
When the noise is of a small intensity, the system 
shows the Poisson behavior. By this reason, in some cases
the neural activity can be described
by Poisson driven ([23, 40] and [8], p.299) stochastic models.

The next proposition is the main statement of this subsection.
It is simply a reformulation of some results (Theorem 9 and
Eq.~(30)) from Section 3.
\begin{guess2}
The system, conjugate (in "average" sense) to Wiener neuron
is again a Wiener neuron. The system, conjugate to Poisson
driven neuron is again an (asymptotically) Poisson driven
neuron.
\end{guess2}

Applying the approach described in Sec.~3.4,
the analogous result might also be stated for 
Ornstein-Uhlenbeck neuron ([8], p.299), that
is defined as a diffusion process with
time dependent coefficients $a$ and $b$
(or $\mu$ and $\sigma$) of some special form.

\subsection{New type of memory in neural networks}

We consider the update equation ([8], pp.119, 230, 930), that
describes the dynamics of neural network consisting
of $n$ McCulloch-Pitts neurons
\begin{equation}
 x_{k}(t+1)=\sigma (h_{k}(t)-\theta_{k}) \quad \mbox{where}
 \quad h_{k}(t) = \sum_{j\neq k}w_{k,j}x_{j}(t)
\end{equation}
Here $w_{k,j}$ are synaptic strengths, $\theta_{k}$ are
threshold constants, $\sigma$ is activation function,
$h_{k}$ is synaptic potential, variable $x_{k}$ stands
for $k$-th neuron binary states, and $t$ designates
the discrete time. Different choices of $\sigma$,  
so-called sigmoid functions, are possible.
It is accepted to include the probabilistic "noisy" term to
this equation ([8], p.930),
\begin{equation}
 P(x_{k}(t+1) = \delta) =
 \frac{1}{2}(1+(-1)^{\delta}\pi_{k}) \ ;
\end{equation}
here, it can be chosen, e.g. ([8], p.902)
 $$\pi_{k}=tanh [T^{-1}h_{k}(t)] $$
where the variable $T>0$ (temperature) reflects
the level of noise.

The main two problems, investigating in neural networks are
retrieval and learning problems. The first one studies the dynamics
of neural states $x_{k}$ provided the connections $w_{i,j}$ are
time independent and fixed. The most interest consists in
revealing the attractors of dynamics (54), which are considered as
the memory storage of given network. The basic result is
the Hopfield theorem, establishing that under some restrictions on
matrix $W$ the configuration point
\begin{equation}
 {\bar x} = {\bar x}(t) =
(x_{1}(t), \ x_{2}(t), \ldots , x_{n}(t), \ldots)
\end{equation}
converges to some fixed-point attractors.
The Hopfield nets consist of spin neurons,
taking values $\pm 1$. It is proposed [28] the existence
of some 'energy' function ([8], p.363) associated with $W$
\begin{equation}
  E = E(W) =
  - \frac{1}{2}\sum_{i\neq j} w_{i,j}s_{i}s_{j}
\end{equation}
which, provided certain restrictions, permits
the Lyapunov function ([8], pp.363, 230):
the value of $E$ is increased
with any update of spins. The local minima points of the function
$E$ are treated as the attractor memory states. This memory
can take about $20\%$ of the whole configuration space [28].
Some other works in neural networks ([8], p.258), generalizing
the Hopfield's approach, introduce
and study
the networks of chaotic elements. The aim is to
provide
the existence of an hierarchical memory storage of
coexisting attractors. 

The analysis from previous sections reveals a new type
of attractors of the dynamics (54) and therefore, a new type
of memory in neural networks. Indeed, for arbitrary $k$-th 
neuron we consider its states 
 $$
  {\bar\xi}_{k} =
 (x_{k}(1),\ x_{k}(2), \ldots, x_{k}(t), \ldots)
 $$
defined by (55), as a discrete random process.
Despite of we were mostly dealt with independent random variables,
the results of Section 3.3 show that the difference analysis is also applicable to 
the case of binary random Markov processes - an important 
restriction (see e.g., [8, 27, 28]), usually imposing  on the
process (53) (or (54)) of the brain states. 
The relation (26) in principle allows to compute all 
of the
final processes ${\bar\xi}_{k}^{(\infty)}(\Lambda)$,
corresponding to those $\Lambda \subset N$, for which the final
probabilities $\pi^{(\infty)}(\Lambda)$ exist. The discrete
processes
\begin{equation}
      ({\bar\xi}_{1}^{(\infty)}(\Lambda_{1}), \
      {\bar\xi}_{2}^{(\infty)}(\Lambda_{2}), \ldots
      {\bar\xi}_{n}^{(\infty)}(\Lambda_{n}),\ldots )
\end{equation}
can be treated as some final (or stationary) processes of the
network dynamics, given by Eqs.~(54) and (55). The 
Cartesian products
\begin{equation}
 {\cal A}_{s_{1}}^{(1)}\times {\cal A}_{s_{2}}^{(2)}\times
 \cdots \times {\cal A}_{s_{k}}^{(k)} \times \cdots
\end{equation}
$(1\leq s_{i} \leq \infty)$ where
${\cal A}^{(k)}_{n}$ is an attractor for the conjugate
orbit, on which the random variable
${\bar\xi}_{k}^{(\infty)}(\Lambda_{n})$ is localized,
are the attractors for configuration point (55) and 
therefore
can be treated as a new type of memory of a given 
neural network.

The learning or task-adapted problems, considering in neural
networks theory, have an inverse statement: given set of patterns
of configuration points to determine the matrix $W$ for
which the fixed point attractors coincide with this set of
patterns. It is supposed
that some finite number of given patterns to be
learned by network, as well as their components, are random and
have identically distributed components ([8], p.651).
In the framework of the analysis presented, the learning problem
can be stated as follows: given  random processes of the form
\begin{equation}
{\bar \eta} = (\eta_{1}, \eta_{2}, \ldots, \eta_{n}, \ldots)
\end{equation}
to determine the matrix $W$ of adjustable connections in such
a way that these $\bar\eta$ coincide with some final
processes (58) with some $\Lambda_{i}\subset N$.

\subsection{Elimination of noise}

The real numbers are sometimes represented in the "pulse density
system" ([31]) -- e.g., the limit in the relation (36) represents
the number $p$ in this system.
The density of finite number of alternating signals 
is easily expressed through the densities of its constituents. 
In this system, all the basic 
operations with signals are also possible ([31]). This 
representation is also used on
the studying the multidimensional stochastic systems 
(e.g., on investigating the baker's transformation [33]).
The difference attractor $\cal A$ for such complex system 
reflects
the synthetic information on all the constituents. In respect
of information alteration, due to the timing of the signals, 
see also [8], p.693.

We have (see Eqs.~(31) and (33))  
$$
\gamma ({\bar\xi}, \Lambda) = \lim_{k\to\infty, k\in \Lambda}
\frac{1}{k}\sum_{i=0}^{k-1}\xi_{i}^{(k)}
$$
and therefore, the $\gamma$ is a type of density. It is 
mentioned in [31]
a device (the charge capacitor), transforming the pulse
density to some analog quantity. This remark
in fact indicates a way to set a correspondence between the 
$\gamma$-coefficient of neural activity and neuron's electrical characteristics.
Hence, the $\gamma$ can be treated as some
mixed analog-digital characteristic of neural activity --
according to von Neumann [31], the neuron enables to combine
the analog and digital features in its activity.

We use these remarks to deduce a new type of 
neuron firing condition, based on analysis from Sec.~3 and
formulated in probabilistic terms.
Accepted condition for neuron firing says that
the total electrical charge in neuron should exceed some
threshold level. 
In the formal neural
networks, it is reflected in the update equation (53) 
-- a weighted sum of input signals should exceeds some threshold
level. 
On the other hand, the work [31]
claims that the actual firing conditions may have a very
different form. 
The next suggestion is derived from consideration of discrete 
random processes. One can see, it has the theoretical-probabilistic 
character and does not refer to any neural context.

Let us have a
neuron, receiving on its input the signals from other neurons. As
in the previous section, we assume that these signals are
some independent random variables $\eta_{k}$, taking the values
$+1$ (excitatory signal) and $-1$ (inhibitory signal).
We assume also that we deal with discrete-time process and that the
$\eta_{k}$ are ordered by the growth of their arrivals time.
Letting $\xi_{k} = (1+\eta_{k})/2$, we have
on the input of neuron a random binary process $\bar\xi$. 
Such processes have been 
considered in Sec.~3 and therefore all the results of this section
can be applied. We are interested in the
statement of Theorem 5, which we now reformulate as follows:
\begin{guess2}
Let $\bar\xi = (\xi_{n})_{n=1}^{\infty}$ be a random binary 
sequence with independent terms $\xi_{n}$,
satisfying (34). If
$$\sum_{n=0}^{\infty}P(\xi_{n}=1) < \infty 
\quad \mbox{then with probability 1}\quad 
H(\gamma\;({\bar\xi})) = 0 \ , $$
and if
$$\sum_{n=0}^{\infty}P(\xi_{n}=1) = \infty \quad \mbox{then with
probability 1}
\quad H(\gamma\;({\bar\xi})) = 1 \ . $$
\end{guess2}

Let us now have infinite number of some binary random variables
(neurons) $\xi_{k}(t)$ with given distribution of probabilities
\begin{equation}
P(\xi_{k}(t) = 0) = p_{k}(t),  \quad
 P(\xi_{k}(t) = 1) = q_{k}(t)
 \qquad (p_{k}(t)+ q_{k}(t) =1) \ .
\end{equation}
Here $k\geq 1$ is neuron's number and $t=0,1,2,\ldots$ is the 
discrete
time. If we understood the realization of the random variable $\xi_{k}(t)$ as
$$
   \xi_{k}(t) =  \left \{ \begin{array}{ll}
   1 & \mbox{k-th neuron is active (fired) at moment t}\\
   0 & \mbox{k-th neuron is inactive (silent) at moment t ,}\\
   \end{array} \right.  
$$
it can be said that every probabilistic neural net is a 
random process ${\cal N}_{P}$ of the form
\begin{equation}
{\bar\xi}(t) = (\xi_{1}(t), \xi_{2}(t), \ldots, \xi_{n}(t),
\ldots)
\end{equation}
Now we introduce a deterministic network ${\cal N}_{D}$,
associated with the process ${\cal N}_{P}$. 
Given 2 infinite matrices: the stochastic matrix
$P=(p_{k}(t))_{k,t=0}^{\infty}$ of the probabilities (60),
and the binary matrix of the connections
$W=(w_{i,j})_{i,j=0}^{\infty}$:
$$
  w_{i,j} = \left \{  \begin{array}{ll}
  1 & \mbox{i-th neuron affects on j-th neuron} \\
  0 & \mbox{i-th neuron does not affect on j-th neuron} \\
  \end{array} \right.
$$
we assign the evolution equation of the net ${\cal N}_{D}$
as follows: 
It is clear that the process
$$ {\bar\xi}(k;t) = (w_{k,1}\xi_{1}(t),
w_{k,2}\xi_{2}(t),  \ldots, w_{k,n}\xi_{n}(t), \ldots )$$
consisting of all the neurons, affecting on $k$-th at the moment
$t$, is the input process for $k$-th neuron.
According to Corollary 6 the quantity
$$\Gamma ({\bar\xi}(k;t)) = H(\gamma ({\bar\xi}(k;t))) $$
($ \Gamma = Ho{\gamma}$ is the composition of
$H$ and $\gamma$) is either $0$ or $1$.
We impose the following firing condition for the neurons of
${\cal N}_{D}$:
\begin{equation}
 x_{k}(t+1) = \Gamma ({\bar\xi}(k;t)) \ .
\end{equation}

Let us explain these definitions.
The equation (62) is the evolution equation of the process
${\bar x}(t)$: given $\bar\xi (t)$ it allows to compute
the ${\bar x}(t+1)$.
This means we have required the $\gamma$-characteristic 
(and hence the entropy $H(\gamma)$)
to be the basic quantity, determining the dynamics of the 
deterministic network ${\cal N}_{D}$:
in order to determine its state at next time step (i.e.
${\bar x}(t+1)$), each neuron of ${\cal N}_{P}$ computes
the $\gamma$-characteristic and then the entropy $H(\gamma)$ 
of its present input (the ${\bar\xi}(k;t)$).

It follows from Corollary 6,
that the dynamics of the net ${\cal N}_{D}$
can be represented in the form 
\begin{equation}
 x_{k}(t+1)=\Delta (S_{k}(t)) \quad \mbox{where}
 \quad S_{k}(t) = \sum_{j\neq k}w_{k,j}x_{j}(t)p_{j}(t)
\end{equation}
(cp.~Eq.~(53)) where $\Delta$ is the impulse function of the form
 $$
 \Delta (x) =   \left \{ \begin{array}{ll}
      1 & x = \infty \\
      0 & x \neq \infty \ .
 \end{array} \right.
 $$

In other words, we have proved the following
\begin{guess1}
For every neural network ${\cal N}_{P}$,
consisting of probabilistic
neurons $\xi_{k}=\xi_{k}(t)$ there exists
some deterministic neural network
${\cal N}_{D}$ which at each time $t$ computes
the $\gamma$-characteristic of the inputs of $\xi_{k}$.
\end{guess1}

The probabilistic "noisy" networks has been introduced by 
W.~Little [27]
in order to include to the theoretical studies the noise actually
presenting in the brain. The Theorem 10 shows, that when we are
interested in the entropy aspects of neural activity and if we  
follow the difference approach, the
noise can be eliminated. 
In this respect, the latter theorem 
can also be treated as some analogy 
of de Leeuw-Moore-Shannon-Shapiro statements [39]
from the automata theory for the case of neural networks. 
Note also that Theorem 10 makes the modified neural networks 
remarkably closer to actual brain: the brain enables to operate  
avoiding the influence of external noise.


\newpage
\begin{thebibliography}{99}
\bibitem[1]{av} A. Yu. Shahverdian and A. V. Apkarian,
{\em Fractals}, 7, 1, 1999.
\bibitem[2]{fr} A. Yu. Shahverdian, {\em Fractals},
8, 1, 2000.
\bibitem[3]{vv} A. Yu. Shahverdian and A. V. Apkarian,
Periodic response of periodically perturbed
stochastic systems, LANL preprint, 2000
\bibitem[4]{ja} A. J. Lichtenberg and M. A. Lieberman,
{\em Regular and Stochastic Motion}, Springer, N-Y, 1983.
\bibitem[5]{pr} I. Prigogine
{\em From Being to Becoming}, Freeman, San Francisco, 1980.
\bibitem[6]{kl} Yu. L. Klimontovich,
{The Turbulent Motion and Structure of Chaos}, Nauka,
Moscow, 1990 (in Russian).
\bibitem[7]{bi} P. Billingsley
{\em The Ergodic Theory and Information}, Wiley, New York, 1965.
\bibitem[8]{gl}
{\em The Handbook of Brain Theory and Neural Networks},
ed. M. A. Arbib, The MIT Press, Massachusetts, 1995.
\bibitem[9]{ri} J. Riordan
{\em Combinatorial Identities}, Wiley, N-Y, 1968
\bibitem[10]{aa} G. E. Andrews
{\em The Theory of Partitions}, Addison-Wesley,
Massachusetts, 1976
\bibitem[11]{zs} Z. Star
{\em Aequations Math}, 1976
\bibitem[12]{mn} F. C. Moon,
{\em Chaotic Vibrations}, Wiley, 1987.
\bibitem[13]{vb} V. Barger and M. Olsson,
{\em Classical Mechanics: Modern Perspective}, 2-nd ed.,
McGraw-Hill, 1995.
\bibitem[14]{hs} H. Schuster,
{\em Deterministic Chaos}, Springer, 1984.
\bibitem[15]{uk} U. Quasthoft
{\em in: Fractals in Physics. Proc. VI Trieste Int.
Symp. on Fractals in Physics, ICTP, Trieste, Italy, 1985.
ed. L. Pietronero and E. Tosatti. 1986}
\bibitem[16]{wf} W. Feller
{\em An Introduction to Probability Theory and its
Applications}, N-Y, 1950.
\bibitem[17]{ml} G. Marsaglia
{\em Ann. Math. Stat.} 42, 6, 1971.
\bibitem[18]{hr} G. H. Hardy
{\em Divergent Series}, Oxford, 1949
\bibitem[19]{pr} Yu. V. Prokhorov and Yu. A. Rozanov
{\em Theory of Probabilities},
Moscow, Nauka, 1987
\bibitem[20]{ba} G. P. Basharin, P. P. Bocharov,
Ya. A. Kogan
{\em Queuing Analysis for Computer Networks. Theory and
Computational Methods.} Moscow, Nauka, 1989.
\bibitem[21]{ht} E. G. Jr. Coffman and M. I. Reiman
{\em Diffusion approximations for computer and communications
systems} in: Math Computer Performance and Reliability,
eds. G. Iazeolla, P. J. Courtois, A. Hordijk,
Amsterdam: North-Holland, 1984, p.33-53.
\bibitem[22]{tp} A. N. Shiryaev
{\em The Probability}, Moscow, Nauka, 1980
\bibitem[23]{bm} G. L. Gerstein and B. Mandelbrot
{\em Random walk models for the spike activity of a single
neuron}, Biophys. J., 4: 41-68, 1964
\bibitem[24]{ms} K. Weisenfeld and  F. Moss,
{\em Nature}, 373, 33-36, 1995.
\bibitem[25]{mf} F. M. Moss, A. Bulsara, M. F. Shlesinger
{\em J. Stat. Phys., 1/2, Proc. NATO Advanced Research
Workshop on Stochastic Resonance in Physics and Biology}, 1993.
\bibitem[26]{vd} D. R. Chialvo and A. V. Apkarian
{\em J. Stat. Phys.}, 70, 375-391, 1993.
\bibitem[27]{wl} W. A. Little
{\em Math. Biosciences}, 19, 1974, 101-120
\bibitem[28]{hh} J. Hopfield
{\em Proc. Natl. Ac. Sci.} 79:2554-2558, 1982
\bibitem[29]{ff} A. Beghdadi, C. Andraud et al.
{\em Entropic and multifractal analysis of disordered
morphologies}, Fractals, 1, 3, 1993.
\bibitem[30]{al} S. Albeverio, J. E. Fenstad,
R. Hoegh-Krohn, T. Lindstrom
{\em Nonstandard Methods in Stochastic Analysis and
Mathematical Physics}, Acad. Press, London, 1986.
\bibitem[31]{jn} J.~von Neumann
{\em The Computer and The Brain}, New Haven, Yale Univ. Press,
1958.
\bibitem[32]{ce} C. E. Shannon
{\em Communication in the presence of noise},
Proc. IEEE, v.72, N 9, 1984
\bibitem[33]{me} N.F.G. Martin and J.W. England 
{\em Mathematical Theory of Entropy}, Cambridge Univ. Press, 
1984
\bibitem[34]{ln} A. Yu. Shahverdian
LANL preprint, 1999
\bibitem[35]{ds} A. Yu. Shahverdian,
Russ. Math. Surveys, v. 47, 1992
\bibitem[36]{sp} F. Reike, D. Warland, R. de Ruyter van
Steveninck, W. Bialek
{\em Spikes. Exploring the Neural Code},
The MIT Press, Massachusetts, 1997
\bibitem[37]{hm} N. Chomsky and G. A. Miller
{\em Finitary models of language users}, in:
Handbook of Mathematical Psychology, vol. 2, Wiley, N-Y, 1963.
\bibitem[38]{sy} Symposium. The design of machines to simulate
the behavior of the human brain, Trans. IRE, EC-5, N 4, 1956.
\bibitem[39]{ss} {\em Automata Studies},
ed. C. E. Shannon and J. McCarthy, Princeton Univ. Press, 1956.
\bibitem[40]{mg} L. Glass and M. Mackey
{\em From Clocks to Chaos. The Rhythms of Life}
Princeton Univ. Press, 1988
\bibitem[41]{ny} G. Gouesbet et al.
{\em Ann. N-Y Ac. Sci.}, v.808, 1997. 
\end{thebibliography}
\end{document}